\documentstyle[multicol,aps,prb,epsf]{revtex}
\newcommand{\Vec}[1]{\mbox{\boldmath$#1$}}
\begin{document}

\draft

\title{The spin-triplet superconductivity induced by the charge
fluctuation in extended Hubbard model
}

\author{Seiichiro Onari$^{1*}$, Ryotaro Arita$^{1\dagger}$, 
Kazuhiko Kuroki$^2$, and Hideo Aoki$^1$}

\address{$^1$Department of Physics, University of Tokyo, Hongo,
Tokyo 113-0033, Japan}
\address{$^2$Department of Applied Physics and Chemistry,
University of Electro-Communications, Chofu, Tokyo 182-8585, Japan}

\date{\today}

\maketitle
\begin{abstract}
The pairing symmetry in the electron mechanism for superconductivity 
is explored when charge fluctuations coexist with 
spin fluctuations.  The extended Hubbard model is adopted 
to obtain, with the fluctuation exchange approximation, 
a phase diagram against the on-site Coulomb repulsion $U$ and 
the off-site repulsion $V$ for the square lattice with second-neighbor 
hopping $t'$.  We have found that (i) for large $U(>9)$ 
a triplet superconductivity 
with a $\sin(k_x+k_y)$ symmetry can appear just below the charge density wave 
phase.  The pairing is degenerate with $\sin(k_x-k_y)$, so a chiral
$\sin(k_x+k_y)+i\sin(k_x-k_y)$ that breaks the time reversal symmetry 
should result, which is a 
candidate for the gap function on the $\gamma$ band of Sr$_2$RuO$_4$ 
and is consistent with a recent measurement 
of the specific heat.  (ii) By systematically 
deforming the Fermi surface with varied $t'$, we have 
identified 
the region where the triplet pairing is most favored 
to be the region where the Fermi surface 
traverses the van Hove singularity with the charge 
susceptibility strongly enhanced.

\end{abstract}


\begin{multicols}{2}

\section{INTRODUCTION}
Spin-triplet superconductivity, which is arousing much interests in 
recent years, is fascinating in a number of ways.  
Theoretically, a most intriguing question is the following: 
if we consider the electron mechanism of superconductivity 
in the most frequently adopted Hubbard model with an on-site 
electron-electron repulsion, we can show that triplet 
superconductivity is very difficult to realize for the 
simple reason that the pairing interaction mediated by 
spin fluctuations is only 1/3 in the triplet channel than 
in the singlet channel\cite{arita2,monthoux}.  
So any theory attempting to explain triplet pairing 
has to overcome this question.  

Experimentally, the discovery of superconductivity by 
Maeno and coworkers in the
layered perovskite ruthenium oxide Sr$_2$RuO$_4$ have kicked off 
renewed interests.  Suggestions for a triplet pairing in this material 
came from NMR Knight shift\cite{ishida1,ishida2} and polarized neutron
scattering\cite{duffy}.  A broken time reversal symmetry is 
further observed with 
$\mu$SR\cite{luke1,luke2} and small-angle neutron scattering\cite{kealey}.
As for the pairing symmetry, NMR and NQR relaxation rates have shown 
an absence of the Hebel-Slichter
peak\cite{ishida3,ishida4}, which suggests line node(s) in the gap 
function, which was supported by a specific heat measurement\cite{nishizaki}.
While the angular dependence of thermal conductivity indicates the presence of
horizontal line-nodes\cite{izawa}, 
a recent field-orientation dependence of the specific heat shows
that the gap in the active $\gamma$ band has minima along [100] directions 
with the passive $\alpha$ and $\beta$ bands having gap minima 
along [110] directions\cite{deguchi}.

A theoretical work by Zhitomirsky and Rice\cite{zhitomirsky}
has proposed horizontal line nodes with a good fit to the specific
heat measurement\cite{nishizaki}.   
As for the mechanism that stabilizes triplet pairing, Kuwabara and
Ogata\cite{kuwabara},
and independently Sato and Kohmoto\cite{sato}, have suggested that triplet $p$-wave
pairing can be induced by an anisotropy in the spin fluctuation.  Kuwabara
and Ogata\cite{kuwabara} have identified the most suitable
gap function to be $\sin k_x+i\sin k_y$ that breaks the 
time-reversal symmetry, which was also
proposed by Miyake and Narikiyo\cite{miyake}.  
The chiral
$\phi_x(k)+i\phi_y(k)$ ($\phi_x(k), \phi_y(k)$: $p$-wave
like) has also been obtained with a third-order
perturbation theory by 
Nomura and Yamada\cite{nomura1,nomura2} and more 
recently by Yanase and Ogata\cite{yanase}, and 
with the fluctuation exchange approximation (FLEX) for quasi-one-dimensional
$\alpha$ and $\beta$ bands by  Kuroki {\it et al.}\cite{kuroki}.  

However, the validity of perturbation results 
truncated at finite orders has to be checked by non-perturbative methods.
Indeed, a recent quantum Monte Carlo 
(QMC) result by Kuroki {\it et al.} 
have shown that the singlet $d$-pairing dominates over the triplet 
for the $\gamma$ band.\cite{kuroki2}
This situation has led Arita {\it et al.} to show, 
with the dynamical cluster approximation (DCA), that
$d_{x^2-y^2}$ pairing actually dominates over $p$-wave pairings 
as far as the on-site Hubbard model is concerned, while if we 
go over to the extended Hubbard model with off-site repulsion 
the triplet superconductivity
can be favored but not dominant in extended Hubbard model employing
nearest neighbor Coulomb repulsion for $\gamma$ band in Sr$_2$RuO$_4$.\cite{arita}

Generally, there is a FLEX result by Arita et al\cite{arita2} who show that 
triplet superconductivity is much weaker than singlet pairs 
for the one-band, on-site Hubbard
model, which agrees with a phenomenology\cite{monthoux}, while 
if charge fluctuations are enhanced by off-site repulsions 
triplet pairs have a chance to dominate.  
We have already shown triplet superconductivity can be dominant near the
charge density wave (CDW) phase on
square lattice with the FLEX in the extended Hubbard model.\cite{onari}

So the purpose of this paper is to examine how triplet 
superconductivity can become dominant in the extended Hubbard model with
the FLEX in general, and for $\gamma$ band of Sr$_2$RuO$_4$ in particular.  
The extended Hubbard
model has been studied, primarily for specific charge densities $n$, 
e.g., half-filling or quarter filling, 
by means of the quantum Monte Carlo method\cite{Zhang}, the weak coupling 
theory\cite{Tesanovic}, the mean-field approximation\cite{Murakami,Seo}, 
the second-order 
perturbation\cite{Onozawa}, the random phase approximation\cite{Scalapino3D},
the FLEX approximation\cite{esirgen3}, 
the slave-boson technique\cite{Merino}, the bosonization and the
renormalization group\cite{Sano,Kuroki-Kusakabe}. 
Here we adopt the FLEX developed by Bickers
{\it et al.}\cite{bickers1,bickers2,dahm,bennemann},
which is a renormalized perturbation method to study pairing instabilities when 
exchange of spin and charge fluctuations are considered as dominant
diagrams. This approximation is useful to explore the tendencies of
dominant pairing relatively when the parameters are systematically varied.
However, including the off-site Coulomb repulsion $V$ susceptibilities
and effective interactions become $(Z+1)\times (Z+1)$ matrices for the
lattice coordination number $Z$ ($=4$ for the square lattice), which
demand extremely computer resources.
We show as far as the present finite-temperature result is concerned
that (i) triplet $\sin(k_x+k_y)$ pairing appears in between
singlet $\cos2k_x-\cos2k_y$ and the CDW phase for $U>9$, and (ii) the
pairing symmetry changes as $d_{x^2-y^2}$ $\rightarrow$ $\sin(k_x+k_y)$
$\rightarrow$ $\cos2k_x-\cos2k_y$ when the shape of the 
Fermi 
surface is varied by the second-neighbor hopping $t'$.  
Physically, all the results can be well explained by the structure 
and peak value of spin and charge
susceptibilities.

\section{FORMULATION}
\begin{figure}[h]
\begin{center}
\leavevmode\epsfysize=30mm 
\epsfbox{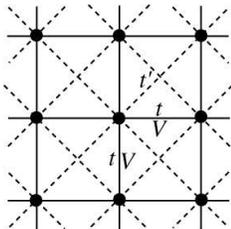}
\caption{A square lattice with nearest-neighbor hopping $t$
 and second neighbor hopping $t'$ 
 with the on-site Coulomb repulsion $U$ and the nearest-neighbor 
repulsion $V$}
\label{lattice}
\end{center}
\end{figure}

Let us start with the extended Hubbard Hamiltonian,
\begin{eqnarray}
{\cal H} = &-&\sum_{i,j}^{\rm nn,nnn}\sum_{\sigma}t_{ij}c_{i\sigma}^{\dagger}c_{j\sigma}+U\sum_{i}n_{i\uparrow}n_{i\downarrow}\\ \nonumber 
&+&\frac{1}{2}\sum_{i,j}^{\rm nn}\sum_{\sigma\sigma'}V_{ij}n_{i\sigma}n_{j\sigma'},
\end{eqnarray}
in the standard notation on a tetragonal lattice depicted in
Fig.\ref{lattice}, where nn (nnn) denotes nearest neighbor
(next-nearest neighbor) sites.
For the square lattice 
the unit of energy is taken to be the nearest-neighbor $t_{ij}=1.0$, and 
lattice constant $a=1$.

To determine the dominant gap function, we solve 
{\'E}liashberg's equation with the FLEX approximation, 
\begin{eqnarray}
\lambda\phi(k)&=&
-\frac{T}{N}\sum_{k'}\Gamma(k,k')G(k')G(-k')\phi(k')
\label{elia},
\end{eqnarray}
where $\phi$ is the gap function, $G$ Green's function, and 
$\Gamma$ the pairing interaction with $k\equiv (\Vec{k},\omega_n)$. 
The eigenvalue $\lambda$, a measure of the pairing, becomes unity 
at $T=T_C$.  For the calculation 
we take an $N=32\times 32$ lattice, the temperature 
$T=0.03$, and the Matsubara frequency for fermions
$-(2N_c-1)\pi T \leq \omega_n \leq (2N_c-1)\pi T$ with $N_c=1024$.

Esirgen {\it et al.}\cite{esirgen3,esirgen1,esirgen2} have extended 
the FLEX method to general lattice Hamiltonians including 
the extended Hubbard model.  Following them we introduce 
the pairing interaction,
{\small
\begin{eqnarray}
\Gamma_{\rm s}(k,k')
=\sum_{\Delta\Vec{r},\Delta\Vec{r}'} \left\{\right.
&\frac{3}{2}&\left[V_{\rm sp}\chi_{\rm sp}V_{\rm sp}\right](k-k';\Delta\Vec{r};\Delta\Vec{r}')e^{i(\Vec{k}\cdot\Delta\Vec{r}+\Vec{k'}\cdot\Delta\Vec{r}')}\nonumber\\
-&\frac{1}{2}&\left[V_{\rm ch}\chi_{\rm ch}V_{\rm ch}\right](k-k';\Delta\Vec{r};\Delta\Vec{r}')e^{i(\Vec{k}\cdot\Delta\Vec{r}+\Vec{k'}\cdot\Delta\Vec{r}')} \nonumber\\
+&\frac{1}{2}&V_{\rm s}(0;\Delta\Vec{r};\Delta\Vec{r}')e^{i(\Vec{k}\cdot\Delta\Vec{r}'-\Vec{k'}\cdot\Delta\Vec{r})}
\left.\right\},
\label{s}
\end{eqnarray}
}
for singlet pairing, and
{\small
\begin{eqnarray}
\Gamma_{\rm t}(k,k')
=\sum_{\Delta\Vec{r},\Delta\Vec{r}'} \left\{\right.
-&\frac{1}{2}&\left[V_{\rm sp}\chi_{\rm sp}V_{\rm sp}\right](k-k';\Delta\Vec{r};\Delta\Vec{r}')e^{i(\Vec{k}\cdot\Delta\Vec{r}+\Vec{k'}\cdot\Delta\Vec{r}')}\nonumber\\
-&\frac{1}{2}&\left[V_{\rm ch}\chi_{\rm ch}V_{\rm ch}\right](k-k';\Delta\Vec{r};\Delta\Vec{r}')e^{i(\Vec{k}\cdot\Delta\Vec{r}+\Vec{k'}\cdot\Delta\Vec{r}')}\nonumber\\
+&\frac{1}{2}&V_{\rm t}(0;\Delta\Vec{r};\Delta\Vec{r}')e^{i(\Vec{k}\cdot\Delta\Vec{r}'-\Vec{k'}\cdot\Delta\Vec{r})} 
\left.\right\}
\label{t}
\end{eqnarray}
}
for triplet pairing.  Here $\Delta\Vec{r}(=\Vec{0},\pm\hat{\Vec{x}},\pm\hat{\Vec{y}})$ is 
null or nearest-neighbor vectors, 
\begin{eqnarray}
\chi_{\rm sp} &=& \overline{\chi}/(1+V_{\rm sp}\overline{\chi}), \nonumber\\
\chi_{\rm ch} &=& \overline{\chi}/(1+V_{\rm ch}\overline{\chi}) \nonumber
\end{eqnarray}
are the spin and charge susceptibilities, respectively, where 
$\overline{\chi}$ is the irreducible susceptibility,
\begin{eqnarray}
\overline{\chi}(q;\Delta\Vec{r};\Delta\Vec{r}')=-\frac{T}{N}\sum_{k'}e^{i\Vec{k'}\cdot(\Delta\Vec{r}-\Delta\Vec{r}')}G(k'+q)G(k'),
\end{eqnarray}
and $V_{\rm ch} (V_{\rm sp})$ is the coupling between density (magnetic) 
fluctuations,
{\small
\begin{eqnarray}
&V_{\rm ch}&(q;\Delta\Vec{r};\Delta\Vec{r}')\nonumber\\
&=& 
\left\{
\begin{array}{ll}
U+4[V_x\cos(q_x)+V_y\cos(q_y)], & \Delta\Vec{r}=\Delta\Vec{r}'=\Vec{0},\\
-V_x, & \Delta\Vec{r}=\Delta\Vec{r}'=\pm\hat{\Vec{x}}\\
-V_y, & \Delta\Vec{r}=\Delta\Vec{r}'=\pm\hat{\Vec{y}}\\
\end{array}
\right.\\
&V_{\rm sp}&(q;\Delta\Vec{r};\Delta\Vec{r}')=
\left\{
\begin{array}{ll}
-U, & \Delta\Vec{r}=\Delta\Vec{r}'=\Vec{0},\\
-V_x, & \Delta\Vec{r}=\Delta\Vec{r}'=\pm\hat{\Vec{x}}\\
-V_y, & \Delta\Vec{r}=\Delta\Vec{r}'=\pm\hat{\Vec{y}},\\
\end{array}
\right.
\end{eqnarray}
}
where $q\equiv(\Vec{q},\epsilon_n)$ with $\epsilon_n=2n\pi T$ 
being the Matsubara frequencies for bosons, and $V_x=V_y=V$ here.
We have found that the $q$ dependence 
of $V_{\rm sp}$ and $V_{\rm ch}$ does not significantly 
affect $\Gamma_{\rm s}$ and $\Gamma_{\rm t}$.  
Accordingly the peak position of $\chi_{\rm ch}$ is almost the same as that for 
$V_{\rm ch}\chi_{\rm ch}V_{\rm ch}$ term in the expression for $\Gamma$.

$V_{\rm s}(0;\Delta\Vec{r};\Delta\Vec{r}')$, $V_{\rm t}(0;\Delta\Vec{r};\Delta\Vec{r}')$, appearing in the last lines in eqs.(\ref{s},\ref{t}) respectively, 
are constant terms involving $U$ and $V$,
\begin{eqnarray}
V_{\rm s}(\Vec{q};\Delta\Vec{r};\Delta\Vec{r}')&=&
\left\{
\begin{array}{rl}
2U, & \Delta\Vec{r}=\Delta\Vec{r}'=\Vec{0},\\
V_x, & \Delta\Vec{r}=\Delta\Vec{r}'=\pm\hat{\Vec{x}}\\
V_xe^{\pm iq_x}, &\Delta\Vec{r}=-\Delta\Vec{r}'=\pm\hat{\Vec{x}}\\
V_y, & \Delta\Vec{r}=\Delta\Vec{r}'=\pm\hat{\Vec{y}}\\
V_ye^{\pm iq_y}, &\Delta\Vec{r}=-\Delta\Vec{r}'=\pm\hat{\Vec{y}}\\
\end{array}
\right.\\
V_{\rm t}(\Vec{q};\Delta\Vec{r};\Delta\Vec{r}')&=&
\left\{
\begin{array}{rl}
V_x, & \Delta\Vec{r}=\Delta\Vec{r}'=\pm\hat{\Vec{x}}\\
-V_xe^{\pm iq_x}, &\Delta\Vec{r}=-\Delta\Vec{r}'=\pm\hat{\Vec{x}}\\
V_y, & \Delta\Vec{r}=\Delta\Vec{r}'=\pm\hat{\Vec{y}}\\
-V_ye^{\pm iq_y}, &\Delta\Vec{r}=-\Delta\Vec{r}'=\pm\hat{\Vec{y}}\\
\end{array}
\right.
\end{eqnarray}
 
When the off-site interaction $V$ is introduced 
all the vertices ($V_{\rm sp}$, $V_{\rm ch}$, $V_{\rm s}$, $V_{\rm t}$) 
as well as the susceptibilities become 
$(Z+1)\times (Z+1)$ matrices for the lattice coordination number $Z 
(=4$ for the square lattice).

\section{RESULT}
\subsection{A case study for Sr$_2$RuO$_4$}
\begin{figure}[h]
\begin{center}
\leavevmode\epsfysize=50mm 
\epsfbox{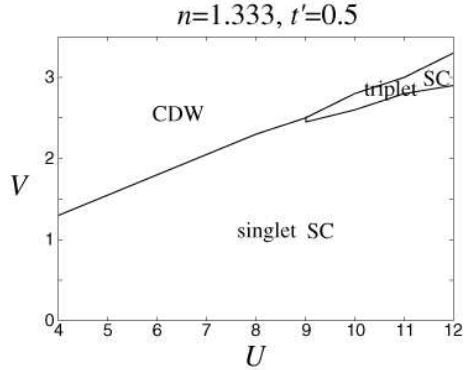}
\caption{Phase diagram against $V$ and $U$ with 
$n=1.333$ and $t'=0.5$ for the 2D extended Hubbard model.  
}
\label{phase}
\end{center}
\end{figure}

We consider $\gamma$ band of Sr$_2$RuO$_4$ whose band filling is $n=4/3$.
To represent the shape of the Fermi surface obtained by ARPES\cite{ARPES}, we
choose parameters $n=1.333$ and next nearest hopping $t'=0.5$. 
We first show the phase diagram against on-site Coulomb interaction
$U$ and the nearest neighbor Coulomb interaction $V$
(Fig. \ref{phase}).  There, we have assumed 
that the dominant superconductivity pairing
is the one that has the largest eigenvalue $\lambda$ in
{\'E}liashberg's equation calculated at $T=0.03$.  
While the value of $\lambda$ for $T=0.03$ is still much smaller than
unity, it is difficult to extend the FLEX calculation to lower
temperatures, given the complexity of the (extended Hubbard) model.  
So this amounts to an assumption that the order in which $\lambda$'s 
appear for various pairing symmetries 
does not change for $T\rightarrow 0$.  
The charge density wave (CDW) is identified as the region in which the
charge susceptibility (which has a peak at $(\pi,\pi)$) diverges at
$T=0.03$.

\begin{figure}[h]
\begin{center}
\leavevmode\epsfysize=110mm 
\epsfbox{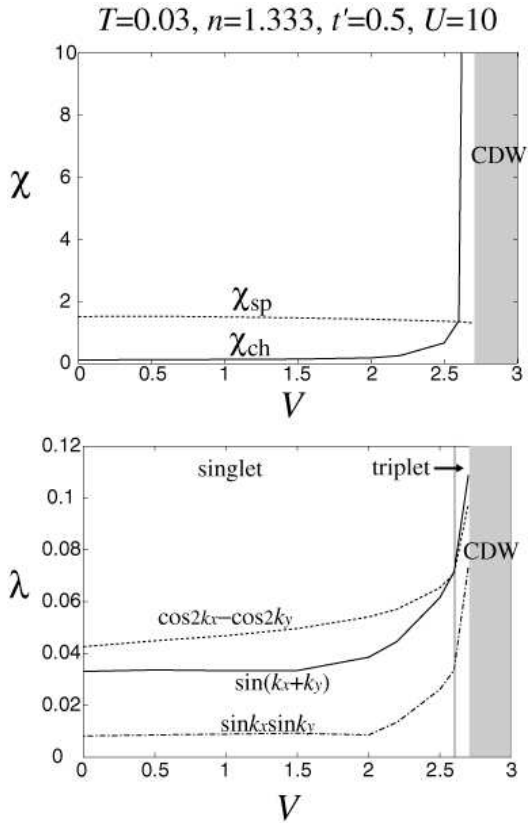}
\caption{Top: the maximum eigenvalue of $\chi_{\rm ch}(\Vec{k},0)$ charge (solid
 line) and $\chi_{\rm sp}(\Vec{k},0)$ spin (dotted)
 susceptibilities as a function of $V$ for $U=10.0$. Bottom: 
 the eigenvalue $\lambda$ of {\'E}liashberg's equation for
 triplet $\sin(k_x+k_y)$ (solid line), singlet $\cos2k_x-\cos2k_y$ (dotted) and
 $\sin k_x+\sin k_y$ (dot-dash) as a function of $V$ for
 $U=10.0$. The CDW (gray) region is identified from the divergence in the charge susceptibility.
}
\label{su-csdw}
\end{center}
\end{figure}

\begin{figure}[h]
\begin{center}
\leavevmode\epsfysize=40mm 
\epsfbox{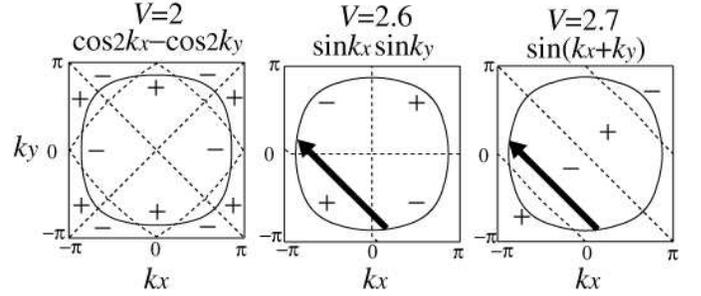}
\caption{The dominant gap function $\sin(k_x+k_y)$ (triplet) in $k$ space for
 $V=2.7$ (right panel), $\cos2k_x-\cos2k_y$ (singlet) for $V=2$ (left), and
 sub-dominant singlet gap function $\sin k_x\sin k_y$ for $V=2.6$ (center),
 each for $U=10.0$. The arrows indicate the main scattering process mediated by
 charge fluctuation.}
\label{gap}
\end{center}
\end{figure}

In the result a triplet superconductivity phase (which does not 
exist for a single band when  $V$ is absent) is seen to appear just below the
CDW phase for $U>9$. 
The maximum eigenvalue of charge and spin susceptibilities and detailed behavior of $\lambda$
as a function of $V$ for fixed
$U=10$ is shown in Fig. \ref{su-csdw}. 
From the figure we see that the triplet superconductivity
becomes dominant when the charge susceptibility $\chi_{\rm ch}$ is
larger than the spin susceptibility $\chi_{\rm sp}$.
Figure \ref{gap} shows gap functions in $k$-space, 
which shows that the triplet (which 
has the highest $\lambda$) has 
a gap function $\propto\sin(k_x+k_y)$), while the two singlet gap functions 
have $\cos2k_x-\cos2k_y$ and $\sin k_x\sin k_y$.\cite{cos}

\begin{figure}[h]
\begin{center}
\leavevmode\epsfysize=40mm 
\epsfbox{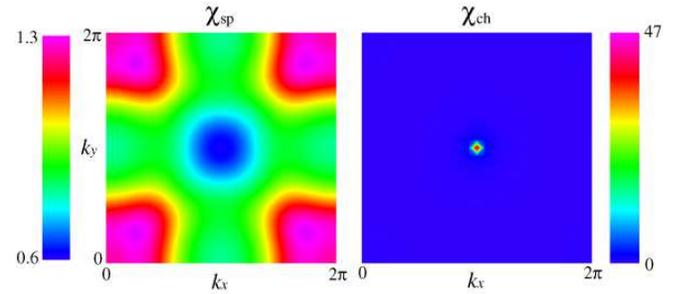}
\caption{Color-coded plots of the spin susceptibility $\chi_{\rm sp}(\Vec{k},0)$ (left) and the
 charge susceptibility $\chi_{\rm ch}(\Vec{k},0) $(right) for $U=10.0, V=2.7$.}
\label{peak}
\end{center}
\end{figure}

We can trace back the reason why these gap functions are favored 
to the structure of spin and charge susceptibilities.
Figure \ref{peak} shows the peak positions of spin and
charge susceptibilities for $V=2.7$. The peak positions for charge are
around $(\pm\pi,\pm\pi),(\pm\pi,\mp\pi)$, while those for spin are
ferromagnetic-like around $(\pm\frac{\pi}{4},\pm\frac{\pi}{4})$,
$(\pm\frac{\pi}{4},\mp\frac{\pi}{4})$.
In {\'E}liashberg's
eq.(\ref{elia})-(\ref{t}), we see that for singlet
the coefficient of $\chi_{\rm sp}$ ($\chi_{\rm ch}$) term in $\Gamma_{\rm
s}$ are positive (negative). So the dominant gap
function must change sign across the momentum transfer 
that points the peak position in the spin susceptibility, while 
the gap function must not change sign across the momentum transfer 
that points the peak position 
in the charge susceptibility. For triplet, on the other
hand, the coefficients of $\chi_{\rm sp}$ and $\chi_{\rm ch}$ terms 
in $\Gamma_{\rm t}$ are both negative. So the gap function should have the same
sign across both the spin and charge peak positions.

When $V$ is small, the spin susceptibility is much greater than 
the charge susceptibility, and the spin-fluctuation 
mediated pairing interaction for 
singlet is three times larger than that for triplet.  This 
explains $\cos2k_x-\cos2k_y$, which changes sign across $(\pm\frac{\pi}{4},\pm\frac{\pi}{4})$,
$(\pm\frac{\pi}{4},\mp\frac{\pi}{4})$, becomes the dominant gap
function.
For large $V$, the charge susceptibility becomes
dominant, and the pairing interaction 
becomes the same between singlet and triplet channels, i.e., 
$\Gamma_s$ and $\Gamma_t$ become similar in magnitude. 
Whether triplet is more favored than
singlet depends on the other factors such as the number of nodes on
the Fermi surface, which work unfavorably for pairing 
since the integration around the node in the right hand side of 
eqn.(\ref{elia}) cannot contribute to
$\lambda$. So a larger the number of nodes is basically unfavored.

As another factor, we can find from Fig. \ref{su-csdw} that the
$\lambda$'s for the gap function $\sin(k_x+k_y)$ and
$\sin k_x\sin k_y$ increase more sharply than that for $\cos2k_x-\cos2k_y$
for $V>2$, where $\chi_{\rm ch}$ increases sharply. The reason should be
that the first two have the same sign across the peak of $\chi_{\rm ch}$
around $(\pm\pi,\pm\pi)$ or
$(\pm\pi,\mp\pi)$, while the latter has slightly warped nodal lines.  
So there is a region exists where the gap function has the 
opposite signs across the peak of $\chi_{\rm ch}$.

\begin{figure}[h]
\begin{center}
\leavevmode\epsfysize=40mm 
\epsfbox{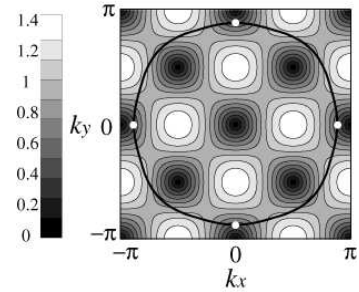}
\caption{Contour plot of $|\sin(k_x+k_y)+i\sin(k_x-k_y)|$ along 
with the Fermi surface for
 $T=0.03, U=10.0, V=2.7$. Open circles indicate minima of the 
 gap function on the Fermi surface.
}
\label{tri}
\end{center}
\end{figure}

Returning to Sr$_2$RuO$_4$, our result in Fig. \ref{su-csdw} indicates that 
if triplet can be
dominant, the gap function should be
$\sin(k_x+k_y)$ which is degenerate with $\sin(k_x-k_y)$. The true gap function
below $T_c$ should be a complex linear
combination, 
\begin{eqnarray}
\sin(k_x+k_y)+i\sin(k_x-k_y),
\end{eqnarray}
which is more stable thermodynamically and breaks the time-reversal
symmetry (Fig. \ref{tri}). The absolute value of the gap function has
minima on the Fermi
surface along $[100]$ and equivalent directions as depicted in
Fig. \ref{tri} with open circles, which is consistent with the recent experiment of the
specific heat in rotated magnetic fields.\cite{deguchi}
However, we also notice that 
the triplet region is very narrow, so identification 
of the pairing in Sr$_2$RuO$_4$ would require a precise 
determination of the parameters in that material.  
In addition, if we include the effect of $\alpha$ and
$\beta$ band of Sr$_2$RuO$_4$ this triplet region may expand.

\subsection{Relation between the pairing symmetry and the Fermi surface}
\begin{figure}[h]
\begin{center}
\leavevmode\epsfysize=50mm 
\epsfbox{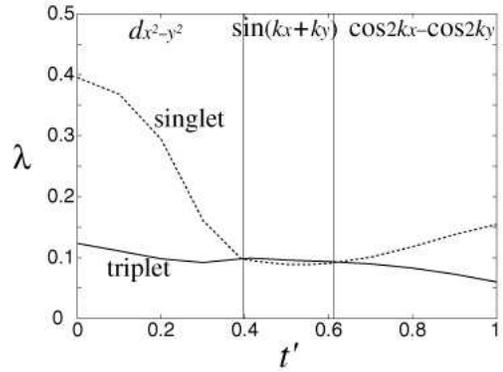}
\caption{The maximum eigenvalue $\lambda$ of {\'E}liashberg's equation for
 spin-triplet (solid line), spin-singlet (dotted) as a
 function of the second neighbor hopping $t'$ for $U=10.0$ and $V=2.7$ 
 with the dominant pairing symmetry indicated}
\label{su-t}
\end{center}
\end{figure}

\begin{figure}[h]
\begin{center}
\leavevmode\epsfysize=80mm 
\epsfbox{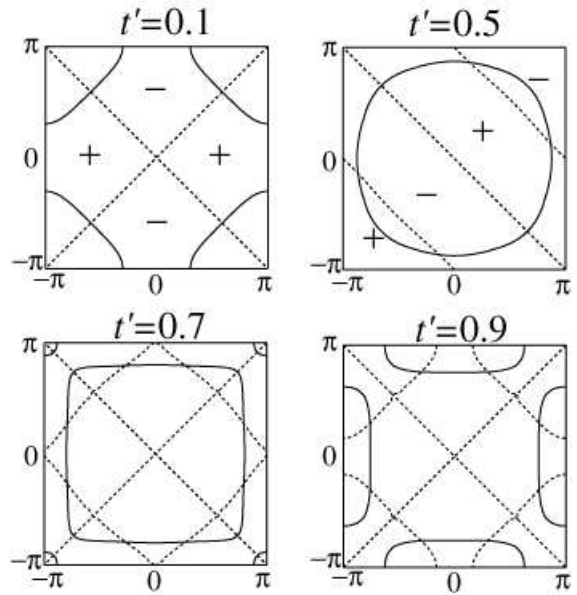}
\caption{Fermi surface (solid line) and nodes of the 
dominant gap function (dotted) for $t'=0.1$ (top left), 
$t'=0.5$ (top right), $t'=0.7$ (bottom left) and $t'=0.9$ (bottom right) 
for $U=10.0$ and $V=2.7$.}
\label{fermi-gap-t}
\end{center}
\end{figure}

\begin{figure}[h]
\begin{center}
\leavevmode\epsfysize=30mm 
\epsfbox{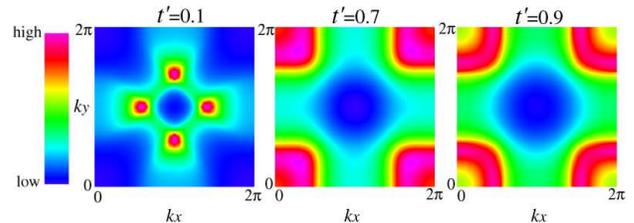}
\caption{Color-coded plots of the spin susceptibility $\chi_{\rm sp}(\Vec{k},0)$  for
 $t'=0.1$ (left), $t'=0.7$ (center), and $t'=0.9$ (right) for
 $U=10.0$ and $V=2.7$.}
\label{sdw-peak}
\end{center}
\end{figure}

\begin{figure}[h]
\begin{center}
\leavevmode\epsfysize=50mm 
\epsfbox{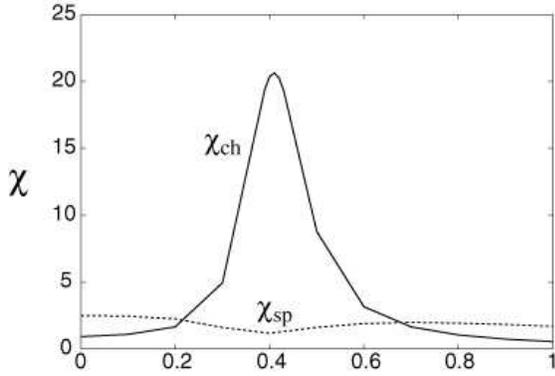}
\caption{ The maximum eigenvalue of $\chi_{\rm ch}(\Vec{k},0)$ 
 (solid line) and $\chi_{\rm sp}(\Vec{k},0)$ (dotted) 
 as a function of $t'$ for $U=10.0$ and $V=2.7$.}
\label{csdw-t}
\end{center}
\end{figure}

Let us identify the relation between the pairing symmetry and
shape of the Fermi surface in the present context.  
For that purpose we can change the second neighbor hopping $t'$, 
which controls the warping (and
even the topology) of the Fermi surface.  
In Fig. \ref{su-t} we show $\lambda$ and the
dominant pairing symmetry when $t'$ 
is varied with fixed $U=10.0$ and $V=2.7$. Figure \ref{fermi-gap-t} depicts
how the Fermi surface obtained by $\epsilon^0_{\bf k}+{\rm Re}\Sigma(k)=\mu$
changes with $t'$. The reason why
those gap functions dominate can be understood in terms of the structure
and value of the spin and charge susceptibilities.
While the peaks in the spin susceptibility for $V=0$ correspond to
the nesting vectors for the Fermi surface, the peaks for large $V$ do
not necessarily correspond to those. The result shows that
the peaks change from $(0.6\pi,\pi)$, $(\pi,0.6\pi)$ for $t'=0-0.3$, $(0.3\pi,0.3\pi)$ for
$t'=0.4-0.8$, to $(0,0.5\pi)$, $(\pi,0.5\pi)$ for $t'=0.9-1.0$ as seen in Fig. \ref{sdw-peak}. On the other hand the
peak position in the charge susceptibility remains at $(\pi,\pi)$ for
the whole range
of $t'$. The peak values of $\chi_{\rm ch}(\Vec{k},0)$ and $\chi_{\rm
sp}(\Vec{k},0)$ are shown in Fig. \ref{csdw-t}, where the peak of $\chi_{\rm
ch}(\Vec{k},0)$ changes rapidly while $\chi_{\rm sp}(\Vec{k},0)$
has an almost constant peak value.

From the result we see that nearly antiferromagnetic spin
fluctuations favor $d_{x^2-y^2}$ pairing for $t'=0-0.3$ as in 
the high-$T_c$ cuprates, strong charge
fluctuations favor $\sin(k_x+k_y)$ for $t'=0.3-0.6$.   For $t'=0.7-1.0$ 
where the
spin susceptibility again exceeds the charge susceptibility, nearly ferromagnetic spin
fluctuations favor $\cos2k_x-\cos2k_y$ which has many sign changes on the
Fermi surface. 

Finally it is interesting to identify the reason why the charge
fluctuation is sharply peaked around $t'=0.4$. 
To elaborate the point, we first note the van-Hove singularities, 
which reside at $(0,\pm\pi)$,
$(\pm\pi,0)$, $(\pm\arccos(-t/(2t')),\pm\arccos(-t/(2t')))$
and $(\pm\arccos(-t/(2t')),\mp\arccos(-t/(2t')))$ for the $t-t'$ tight-binding model.  
So the charge susceptibility with a peak around $(\pi,\pi)$ should be
maximized when the Fermi surface approaches $(0,\pm\pi)$ and $(\pm\pi,0)$, since
the pair hopping with the momentum transfer $(\pi,\pi)$ connects 
the van-Hove singularities, where the density
of state is large. In the terminology of the 
previous work of ours\cite{onari2}, 
the Fermi surface becomes ``thick'' around the
van-Hove singularities.

\begin{figure}[htdp]
\begin{center}
\leavevmode\epsfysize=30mm 
\epsfbox{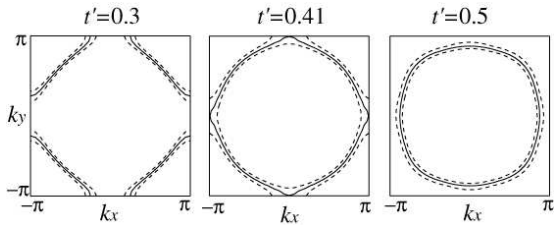}
\caption{The $t'$ dependence of the Fermi surface (solid lines), while dotted
 lines delineate $E=E_F\pm0.2$ for $U=10.0$, $V=2.7$.
}
\label{van}
\end{center}
\end{figure}

We have actually studied the relation between the charge
susceptibility and the Fermi surface as shown in Fig. \ref{van} to confirm
that the charge susceptibility becomes maximum for $t'=0.41$, at which the Fermi
surface just touches the van-Hove singularities $(0,\pm\pi)$,
$(\pm\pi,0)$ and the topology of the Fermi surface changes.
We see in the figure that the Fermi surface becomes ``thick'' around the
van-Hove singularities.  
The value $t'=0.4$ is close to the $t'$ for 
$\gamma$ band of Sr$_2$RuO$_4$, which may imply that triplet
superconductivity induced by charge fluctuations may 
in fact be relevant there.

\subsection{Real-space representation}
\begin{figure}[h]
\begin{center}
\leavevmode\epsfysize=30mm 
\epsfbox{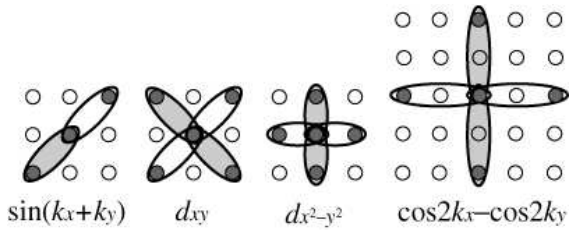}
\caption{ The pairing symmetry in real space.}
\label{real}
\end{center}
\end{figure}

So far we have shown all the results in $\Vec{k}$-space, but it is also
heuristic to represent the pairing symmetry in real space as depicted in Fig. \ref{real}.
From Fig. \ref{su-t}-\ref{csdw-t} we see that $d_{x^2-y^2}$ is
suppressed as the $(\pi,\pi)$ charge fluctuation increases, while from Fig. \ref{su-csdw} $\sin(k_x+k_y)$, $d_{xy}$ and
$\cos2k_x-\cos2k_y$ are favored as the charge fluctuation increases.
This should be because strong $(\pi,\pi)$ charge fluctuations tend 
to arrange electrons
diagonally on the lattice, and suppresses $d_{x^2-y^2}$ pairs across the 
nearest neighbors, but do not disturb the pairing such as 
$\sin(k_x+k_y)$, $d_{xy}$ and
$\cos2k_x-\cos2k_y$ formed across more distant sites.
\section{Conclusion}
We have studied the pairing symmetry in the one-band extended Hubbard
model having the nearest neighbor Coulomb repulsion for square lattice.
For the parameter which represents $\gamma$ band of Sr$_2$RuO$_4$,
spin-triplet superconductivity whose symmetry is $\sin(k_x+k_y)$ appears
just below the CDW phase.  
So the true gap below $T_c$ is suggested to be a chiral
$\sin(k_x+k_y)+i\sin(k_x-k_y)$ which breaks the 
time-reversal symmetry. It would be interesting to experimentally detect
the pairing state using, for example, a theoretical prediction made
in ref. \onlinecite{takigawa}.
However, the region for the spin-triplet superconductivity is very small, 
and the identification of the pairing in Sr$_2$RuO$_4$ will require 
a precise determination of parameters, 
and also a study of the effects of $\alpha$ and
$\beta$ bands. 

We have also examined the relation between the pairing symmetry and
the shape of the Fermi surface by varying $t'$. As $t'$ increases,
the shape of the Fermi surface changes drastically which changes the
structure of $\chi_{\rm sp}$ and $\chi_{\rm ch}$. This in turn changes
the pairing symmetry as $d_{x^2-y^2}\rightarrow\sin(k_x+k_y)\rightarrow\cos2k_x-\cos2k_y$.
In addition, $\chi_{\rm ch}$ becomes especially large when the Fermi surface
touches van-Hove singularities at which the
Fermi surface changes the topology, which is
achieved for $t'=0.41$ near the
parameter for $\gamma$ band of Sr$_2$RuO$_4$.

\section{ACKNOWLEDGMENTS}

Numerical calculations were performed at the supercomputer center, ISSP.

$^*$ Present address: Graduate School of Engineering, Nagoya University,
 Chikusa, Nagoya 464-8603, Japan.\\
$^\dagger$ Present address: Max-Planck-Institut f\"{u}r Festk\"{o}operforschung, Stuttgart, Germany.

\end{multicols}

\begin{references}
\bibitem{cos}In Ref.\cite{kuroki2} the dominant spin singlet gap function
is $d_{x^2-y^2}$ for $U=4$, $t'=0.4$ with $V=0$, where the spin 
structure is antiferromagnetic (peaked at $(\pi,\pi/2)$), 
but our result for $t'=0.5$ has a ferromagnetic-like spin susceptibility 
which suppresses $d_{x^2-y^2}$.
\end{references}
\end{document}